\documentclass[aps,prd,twocolumn,superscriptaddress,nofootinbib,floatfix,noshowpacs]{revtex4-2}
\pdfoutput=1
\newif\ifcomment
\usepackage{bm}
\usepackage{amsmath,amssymb}
\usepackage{color,hyperref,url}
\usepackage{listings}
\usepackage{slashed}
\usepackage[pdftex]{graphicx}
\usepackage{epstopdf}
\usepackage{epsfig}
\usepackage{grffile}
\usepackage{relsize}
\usepackage{float}
\graphicspath{{./img/}}
\usepackage{soul}
\newcommand{\beq}{\begin{equation}}
\newcommand{\eeq}{\end{equation}}
\newcommand{\ba}{\begin{array}}
\newcommand{\ea}{\end{array}}
\newcommand{\bea}{\begin{align}}
\newcommand{\eea}{\end{align}}
\newcommand{\bi}{\begin{itemize}}
\newcommand{\ei}{\end{itemize}}
\newcommand{\ben}{\begin{enumerate}}
\newcommand{\een}{\end{enumerate}}
\newcommand{\bc}{\begin{center}}
\newcommand{\ec}{\end{center}}
\newcommand{\bl}{\begin{flushleft}}
\newcommand{\el}{\end{flushleft}}
\newcommand{\br}{\begin{flushright}}
\newcommand{\er}{\end{flushright}}

\begin{document}
\title{Bridging Electromagnetic and Gravitational Form Factors: Insights from LFHQCD}
\author{Xiaobin Wang}%
\email{wangxiaobin@mail.nankai.edu.cn}
\affiliation{School of Physics, Nankai University, Tianjin 300071, China}
\author{Zanbin Xing}%
\email{xingzb@mail.nankai.edu.cn}
\affiliation{School of Physics, Nankai University, Tianjin 300071, China}
\author{Minghui Ding} \email{m.ding@hzdr.de}
\affiliation{Helmholtz-Zentrum Dresden-Rossendorf, Bautzner Landstra{\ss}e 400, 01328 Dresden, Germany}
\author{Kh\'epani Raya}
\email{khepani.raya@dci.uhu.es}
\affiliation{Department of Integrated Sciences and Center for Advanced Studies in Physics, Mathematics and Computation, University of Huelva, E-21071 Huelva, Spain.}
\author{Lei Chang}%
\email{leichang@nankai.edu.cn}
\affiliation{School of Physics, Nankai University, Tianjin 300071, China}

\begin{abstract}
We propose an efficacious approach to derive the generalized parton distributions for the pion and proton, based upon prior knowledge of their respective parton distribution functions (PDFs). Our method leverages on integral representations of the electromagnetic form factors derived from the light-front holographic QCD (LFHQCD) formalism, coupled with PDFs computed from continuum Schwinger functional methods at the hadronic scale. Using these techniques, we  calculate gravitational form factors and associated mass distributions for each hadron. Remarkably, our calculations yield results that closely match recent lattice QCD simulations conducted near the physical pion mass. This work not only deepens our understanding of hadronic structure but also highlights the efficacy of the LFHQCD approach in modeling fundamental properties of hadrons.
\end{abstract}

\maketitle

\section{Introduction}\label{s1}
The experimental access to the hadronic structure is typically gained via electromagnetic probes, either through exclusive processes that yield electromagnetic form factors (EFFs) or through inelastic experiments that identify parton distribution functions (PDFs)~\cite{Ellis:1996mzs}. These objects have been crucial in exposing the substructure of the proton for the first time and later in revealing various aspects of hadron internal structure, such as charge and momentum distributions. Gravitational form factors (GFFs), on the other hand, provide a complementary perspective by giving access to the mechanical structure of hadrons, thereby exposing, among others, mass and pressure distributions~\cite{Pagels:1966zza,Polyakov:2018zvc}. Extracting GFFs experimentally is more challenging~\cite{Kumano:2017lhr}. However, high-precision lattice QCD (lQCD) data for the pion and proton GFFs have recently become available~\cite{Hackett:2023nkr,Hackett:2023rif}. These analyses were conducted at a pion mass close to its physical value ($m_\pi^{lat}\approx 0.170$ GeV) and also disentangled the contributions from different parton species and flavors. This necessitates significant advancements in our understanding of hadronic internal structure and properties, addressing fundamental aspects such as mass, spin, and $D$-term contributions, which are crucial for unraveling the intricacies of the strong interactions of quantum chromodynamics (QCD). The availability of precise lQCD data thus opens new avenues for further theoretical investigations and experimental validations, especially in this era of high-luminosity and high-energy facilities\,\cite{Accardi:2023chb,Arrington:2021biu,Anderle:2021wcy,Quintans:2022utc,Burkert:2022hjz}.

Perhaps one of the most demanding research avenues is the investigation of generalized parton distributions (GPDs). These quantities offer a comprehensive framework for scrutinizing the structural properties of hadrons\,\cite{Ji:1996ek,Radyushkin:1997ki,Muller:1994ses,Goeke:2001tz,Diehl:2003ny,Belitsky:2005qn,Mezrag:2022pqk}, as they provide a unified description of both the longitudinal momentum and transverse spatial distributions of partons within the hadron\,\cite{Diehl:2002he,Burkardt:2000za}. Notably, GPDs simultaneously encompass PDFs, which are understood as the corresponding forward limit, as well as electromagnetic and gravitational form factors, which can be derived from the lowest 
 and second lowest order Mellin moments. For these reasons, although the empirical determination of GPDs remains a significant challenge, this route could represent a viable alternative to access GFFs in the near future\,\cite{Mezrag:2023nkp}. 

Taking advantage of the connection between GPDs and the aforementioned seemingly simpler distributions, improved parameterizations of the pion GPDs have been developed using limited information on PDFs and/or EFFs, as seen in Refs.\,\cite{Xu:2023bwv,Chavez:2021llq, Chavez:2021koz,DallOlio:2024vjv}. However, a simultaneous analysis with a comparable level of sophistication has not been conducted so far for the proton. Within this line of thought, the novel approach of light-front holographic QCD (LFHQCD) provides a fresh perspective that enables a simultaneous analysis of the structural properties of both the pion and the proton~\cite{Brodsky:2008pf,Brodsky:2007hb,Sufian:2016hwn}. As detailed below, the core idea is that the EFF can be expressed in an integral form, from which the corresponding GPD can be properly identified\,\cite{deTeramond:2018ecg,Chang:2020kjj} and conveniently parameterized in terms of the PDF.

The manuscript is organized as follows: In Section \ref{s2}, we  discuss the general principles of deriving GPDs within the LFHQCD framework, capitalizing on the straightforward identification of GPDs in terms of PDFs. Numerical results on gravitational form factors, as well as the corresponding charge and mass distributions, are presented in Section \ref{s3}. Finally, in Section \ref{s4}, we summarize and discuss the scope of this investigation.

\section{GPDs in the LFHQCD framework}\label{s2}
Within the LFHQCD approach, the electromagnetic form factor $F_\tau(t)$ for arbitrary twist $\tau$ is represented by an Euler Beta function (EBF) as follows~\cite{Brodsky:2007hb,Sufian:2016hwn}:
\begin{eqnarray}
\label{eq:EFFint}
    F_{\tau}(t)&=& \frac{1}{N_{\tau}} B\left(\tau-1,\frac{1}{2}-\frac{t}{4\lambda}\right)\nonumber\\
    &=& \frac{1}{N_{\tau}}\int_0^1(1-y)^{\tau-2}y^{-t/4\lambda-1/2}dy\,,
\end{eqnarray}
where $-t = Q^2$ is the photon momentum transfer square; $N_{\tau}=\Gamma(1/2)\Gamma(\tau-1)/\Gamma(\tau-1/2)$ ensures unit normalization; and $\lambda$ is a mass scale to be determined later. For integer values of $\tau$, the EBF generates a pole structure as:
\begin{equation}
\label{eq:EFFpoles}
    F_\tau(t) \sim \frac{1}{(1-t/M_0^2)(1-t/M_1^2)\cdots (1-t/M_{\tau-2}^2)}\,,
\end{equation}
with the poles located at $M_n^2=4\lambda(n+1/2)$, such that $\sqrt{\lambda}=0.548 \,\text{GeV}=m_\rho/\sqrt{2}$, fixed by the $\rho$ meson mass. 
By considering the leading twist expressions for the pion and proton, $\tau_\pi=2$ and $\tau_p=3$, it is clear that the corresponding EFFs exhibit monopolar and dipolar structures, respectively, manifesting the expected $1/Q^2$ and $1/Q^4$ asymptotic patterns~\cite{Lepage:1980fj}.

Turning back to Eq.~\eqref{eq:EFFint}, through the simple substitution $y=w_{\tau}(x)$, the integral can acquire a seemingly richer structure:
\begin{eqnarray}
    \label{eq:EFFGPD}
    F_\tau(t) &=&\frac{1}{N_{\tau}}\int_0^1dx\,\left[1-w_{\tau}(x)\right]^{\tau-2}\left[w_{\tau}(x)\right]^{-t/4\lambda-1/2}\nonumber\\
    &&\times\frac{\partial w_{\tau}(x)}{\partial x}\;,
\end{eqnarray}
such that, as long as ($x\in[0,1]$)
\begin{equation}
\label{eq:conds}
    w_{\tau}(0)=0,\quad w_{\tau}(1)=1,\quad \frac{\partial w_{\tau}(x)}{\partial x}\geq0\,,
\end{equation}
this change of variables is merely a reparameterization of the EBF, so that the EFF remains invariant and preserves its good properties\,\cite{deTeramond:2018ecg,Chang:2020kjj}; for instance, those captured in Eq.~(\ref{eq:EFFpoles}).

On the other hand, the contribution of a given valence quark with typical flavor $q$ to the total EFF can be read from the zeroth moment of the zero-skewness GPD, $H_v^q(x,\xi=0,t)$\,\footnote{Recall the total EFF is obtained by adding up the individual valence quark contributions, weighted by the corresponding electric charges.}, namely 
\begin{equation}
\label{eq:zerothmoment}
F_\tau^q(t)= \int_0^1dx\, H_v^q(x,\xi=0,t)\,,
\end{equation}
so that, in combination with Eq.\,\eqref{eq:EFFGPD}, one can readily identify the zero-skewness GPD as follows:
\begin{equation}
    H_v^q(x,t):=H_v^q(x,\xi=0,t)=q_{\tau}(x)\left[w_{\tau}(x)\right]^{-t/4\lambda}\,,\label{eq:GPDval}
\end{equation}
where
\begin{equation}\label{eq:qw}
    q_{\tau}(x)=\frac{1}{N_{\tau}}\left[1-w_{\tau}(x)\right]^{\tau-2}\left[w_{\tau}(x)\right]^{-1/2}\frac{\partial w_{\tau}(x)}{\partial x}\,.
\end{equation}

 \begin{figure}[t]
    \centering
    \includegraphics[width=\linewidth]{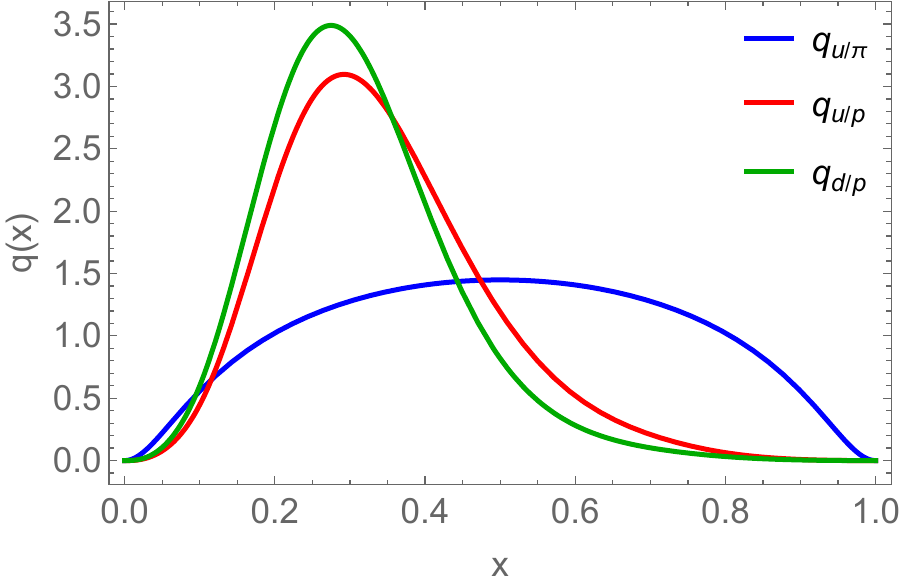}
    \caption{Valence quark PDFs of the pion~\cite{Ding:2019qlr} and proton~\cite{Chang:2022jri} at $\zeta_H$ from CSMs. 
    The PDF of the $u$ quark inside the proton, $q_{u/p}$, has been re-scaled so that it is normalized to 1.
    }
    \label{DFinput}
\end{figure}

The typical approach at this point consists of choosing a profile for the function $w_\tau(x)$ that satisfies the conditions in Eq.~\eqref{eq:conds}. However, the choice of $w_\tau(x)$ is not unique and there are numerous classes of functions that fulfill the necessary requirements (\emph{e.g.} Refs.~\cite{Selyugin:2009ic,deTeramond:2018ecg,Chang:2020kjj}). Here, we propose to reverse this process. That is, with the PDFs rigorously and precisely determined at a certain energy scale $\zeta$, it is possible to use Eq.\,\eqref{eq:qw} to solve the inverse problem of finding $w_{\tau}(x)$. Once this function is determined, via Eqs.~(\ref{eq:GPDval}, \ref{eq:qw}), the zero-skewness GPD $H_v^q(x,t)$ and the distributions that emanate from there can be derived. In other words, the GPD is essentially obtained solely from the prior knowledge of its forward limit. 

We thus consider the valence-quark PDFs obtained using continuum Schwinger functional methods (CSMs) for the pion~\cite{Ding:2019qlr} and the proton~\cite{Chang:2022jri}. These are depicted in Fig.~\ref{DFinput}. The distributions were derived at the so-called hadronic scale $\zeta_H$. This is simply a definition: the resolution scale at which the fully dressed valence quarks fully encode all the properties of the host bound state. However, given the characteristics manifested by the process-independent running coupling in QCD\,\cite{Cui:2019dwv}, as discussed in \emph{e.g.} Ref.\,\cite{Cui:2020tdf}, a value of $\zeta_H\approx 0.331$ GeV can be associated with it. This picture is clearly compatible with considering only the leading-twist components, and many works have successfully adopted these ideas; to name a few, Refs.\,\cite{Xu:2023bwv,Chang:2022jri,Cui:2020tdf,Lu:2022cjx}. Furthermore, the pion and proton PDFs employed herein properly capture crucial aspects of QCD\,\cite{Raya:2024ejx}. Among others, the pion PDF at $\zeta_H$ is a broad and concave function of $x$, with more support than those of the proton; this is a feature of the pion as the Nambu-Goldstone boson\,\cite{Ding:2019qlr}. Moreover, despite considering isospin symmetry ($m_u$ = $m_d$), the valence distributions of the up quark and the down quark within the proton are not proportional to each other. In fact, the calculations do not yield that the up quark carries twice as much proton momentum as the down quark. This fact is attributable to the dynamical formation of diquark correlations within the three-body bound state\,\cite{Lu:2022cjx}. So, we will calculate $w_{3}(x)$ based on the constituent average of the valence distributions, i.e., $q_{3}(x)=(q_{u/p}+q_{d/p})/3$.

\begin{figure}[t]
    \centering
    \includegraphics[width=\linewidth]{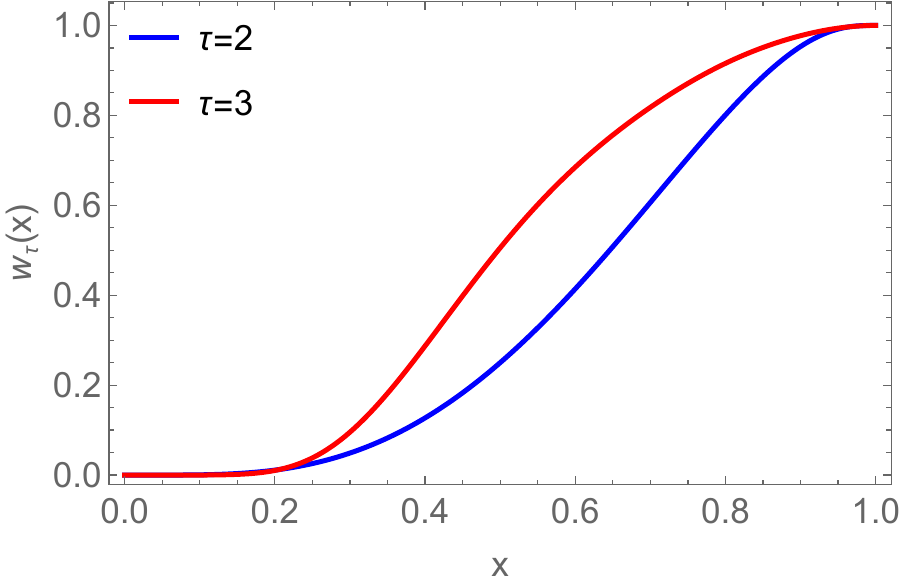}
    \caption{$w_{\tau}(x)$ functions obtained from Eq.~\eqref{eq:qw}, using the input PDFs in Fig.~\ref{DFinput}.}
    \label{winput}
\end{figure}

 The resolution of Eq.\,\eqref{eq:qw} with the CSMs PDFs, as observed in Fig.\,\ref{winput}, produces profile functions $w_{\tau}(x)$ that are plainly compatible with Eq.\,\eqref{eq:conds}. Once more, this guarantees the invariance of the EBF in Eq.~\eqref{eq:EFFGPD}, implying that the EFFs will be the same regardless of the input PDF and will produce the excellent results already known\,\cite{Chang:2020kjj,deTeramond:2018ecg}. Naturally, this shall not be the case for higher order moments nor the GPD itself. The soundness of our selection is then to be tested by calculating the GFFs of the pion and proton associated with the mass, and comparing them with expectations from recent lQCD results\,\cite{Hackett:2023nkr,Hackett:2023rif}. This is discussed in the following section.

\section{Gravitational form factors}\label{s3}
 GFFs provide access to the energy-momentum tensor  in QCD and thus to the mechanical properties of hadrons\,\cite{Polyakov:2018zvc}. Herein, we consider the GFF $A(t)$, which accompanies the $P_{\mu}P_{\nu}$ Lorentz structure for hadrons of any spin and, consequently, is related to the mass distribution inside the hadron. This form factor becomes directly accessible from the first moment of the GPD and, in particular, by taking the zero-skewness limit, the contribution of the $q$-flavored valence quark is straightforwardly obtained as\,\footnote{Note that for the proton we are considering the helicity non-flip GPD.}:
\begin{equation}
    A^{q}(t)=\int_{0}^{1} dx\, x H_{v}^{q}(x,t)\,.
\end{equation}
 The contributions of the valence antiquark and other partons are obtained analogously, and the total form factor results from adding up the contributions from all different parton species. However, when considering the hadronic scale, it is sufficient to consider those coming from the valence quarks since the others would be strictly zero (not so further, since evolution reveals the content of sea quarks and gluons\,\cite{Raya:2021zrz}).

\begin{figure}[t]
    \centering
    \includegraphics[width=\linewidth]{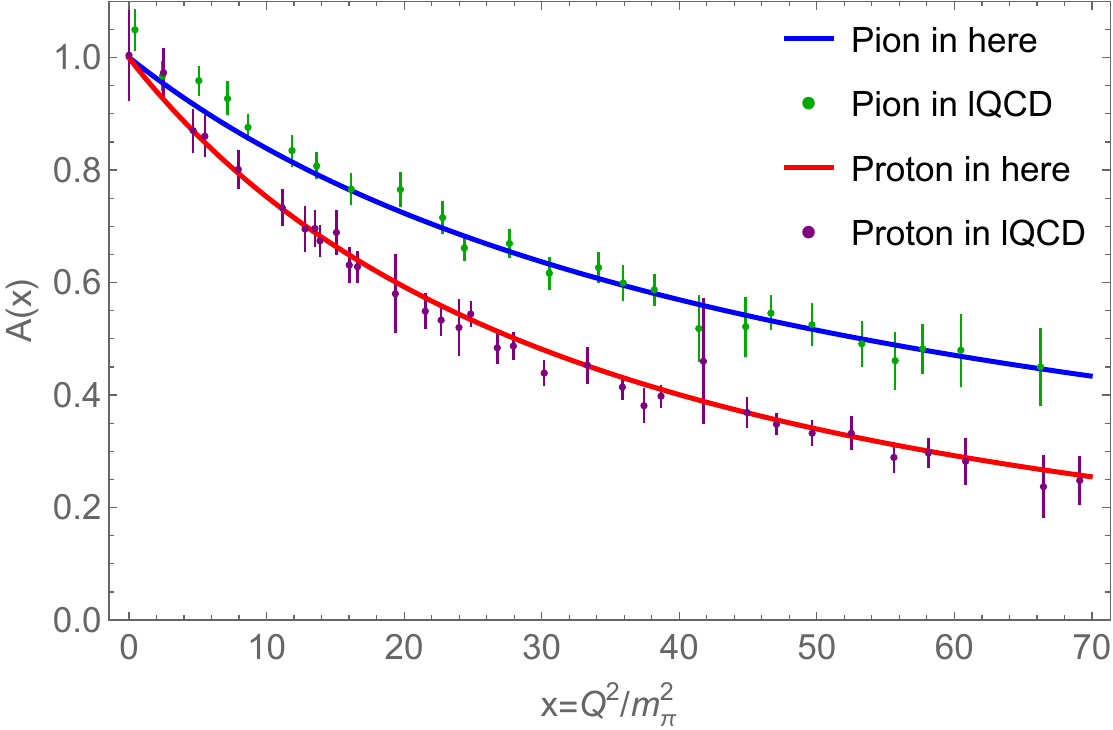}
    \caption{Pion and proton GFFs $A_{\pi,p}(x)$, displayed in terms of $x=Q^{2}/m_{\pi}^2$. LQCD data is taken from Refs.~\cite{Hackett:2023nkr,Hackett:2023rif} ($m_\pi^{here}=0.139$ GeV, $m_\pi^{lQCD}\approx 0.170$ GeV).}
    \label{AFF}
\end{figure}

With the GPDs completely determined as described in the previous section, it is straightforward to calculate the pion and proton GFFs $A_{\pi,p}(t)$. The results are depicted in Fig.~\ref{AFF}, where we also present the corresponding lQCD results from Refs.\,\cite{Hackett:2023nkr,Hackett:2023rif}. It is worth mentioning that the latter have been obtained recently by performing simulations close to the physical mass of the pion, namely $m_\pi^{lat}\approx 0.170$ GeV (our value is $m_\pi=0.139$ GeV). To suppress the effects arising form the slightly higher pion mass, the GFFs in Fig.\,\ref{AFF}, are shown as functions of the dimensionless quantity $x=Q^2/m_\pi^2$.  The results from our present LFHQCD calculation match remarkably well with lQCD expectations for both pion and proton cases.  This alignment not only confirms the effectiveness of the methods employed but also bolsters confidence in the overarching theoretical framework underlying the research, paving the way for explorations with other types of mesons and baryons.

\begin{figure}[t]
    \centering
    \includegraphics[width=\linewidth]{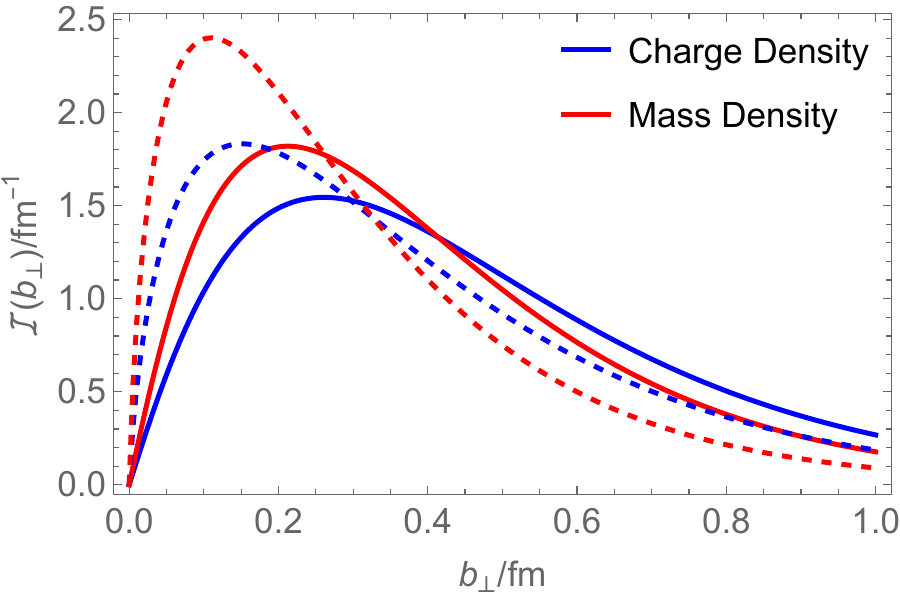}
    \caption{Charge and mass densities $\mathcal{I}^{C,M}(b_\perp)$ in the pion (dashed curves) and proton (solid curves) in the light-front transverse plane.}
    \label{density}
\end{figure}

To provide a more in-depth and intuitive depiction of the internal structure of the hadrons under investigation, let us now consider the charge ($C$) and mass ($M$) densities, $\rho^{C,M}(b_\perp)$, which can be obtained by performing the 2-dimensional Fourier transform of $\mathcal{F}(t) \,(\mathcal{F}=F,A)$:
\begin{eqnarray}
    \rho^{C,M}(b_\perp)&=& \int \frac{d^2Q}{(2\pi)^2}\mathcal{F}(Q^2)e^{-i \bm{Q}\cdot \bm{b_\perp}}\nonumber\\
    &=& \int_0^\infty \frac{dQ}{2\pi}Q J_0(Q b_\perp) \mathcal{F}(Q^2)\,,
\end{eqnarray}
where $J_0$ is the zeroth-order cylindrical Bessel function.
%
%
The computed charge and mass distribution for both the pion and proton are shown in Fig.~\ref{density}. The results reveal that the mass distributions are more localized within a smaller range compared to the charge distributions; and, as expected, the density functions of the proton are broader than those of the pion, indicating that the proton occupies a larger volume, attributed to its greater number of valence components. For the pion, the density
\begin{equation}
    \mathcal{I}^{C,M}(b_\perp):=2\pi b_\perp \rho^{C,M}(|b_\perp|)\,,
\end{equation}
turns out to be maximal at $b_\perp^{max}=0.152,\, 0.11$ fm for the charge and mass distributions, respectively, whereas for the proton, the analogous results yield $b_\perp^{max}=0.261,\,0.213$ fm. The tighter concentration of the mass distributions manifested in both cases suggests that a substantial portion of the hadron's mass is confined within a relatively small region, whereas the larger span of the charge distribution might indicate  combined effects of the individual charged components and their interactions. This hierarchy had already been empirically anticipated in the case of the pion\,\cite{Kumano:2017lhr,Xu:2023bwv}.

Another quantitative measure of the spatial extension acquired by the charge and mass distributions is provided by the corresponding radii, defined as the derivative of the form factors at zero momentum transfer\,\footnote{The difference between calculating $r_{C,M}^2$ in 2 or 3 dimensions is only a factor of $2/3$\,\cite{Kumano:2017lhr}. Typically, the proton's charge radius is deduced from the slope of its charge distribution, encompassing both Dirac and Pauli form factors. However, in the current framework, information on the Pauli term is lacking. Consequently, we estimate the proton's charge radius using the formula $\bar{r}_C^2=r_{C}^2+\frac{6}{4m_{p}^{2}}\chi_{p}$, with $m_{p}=0.94$~GeV and $\chi_{p}=1.793$ representing the proton's mass and anomalous magnetic moment, respectively.}
\emph{i.e.}, $r_{C,M}^2 = -6 \mathcal{F}'(0)$.
%
%
\begin{figure}[t]
    \centering
    \includegraphics[width=\linewidth]{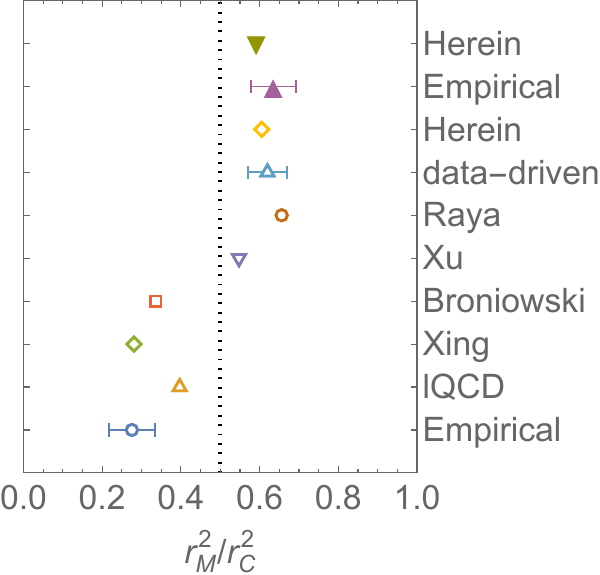}
    \caption{An assortment of evaluations on the ratio  $(r_M/r_C)^2$ in the pion, including (from bottom to top): empirical extractions\,\cite{Kumano:2017lhr},  lQCD results\,\cite{Hackett:2023nkr}, the contact interaction model by Xing\,\cite{Xing:2022mvk}, recent meson dominance analysis by Broniowski\,\cite{Broniowski:2024oyk}, Dyson-Schwinger calculations by Xu\,\cite{Xu:2023izo}, the value inferred via an algebraic GPD ansatz by Raya\,\cite{Raya:2021zrz}, data-driven statistical estimations\,\cite{Xu:2023bwv}, and the result derived herein. Pion ratios are represented by open markers, and for the sake of comparison, the ratios obtained for the proton in this work and empirical extractions\,\cite{Wang:2021dis} are also included in the two uppermost positions, presented as solid markers.}
    \label{radius}
\end{figure}
The estimations derived herein produce a charge-mass squared ratio:
\begin{equation}
    \left(\frac{r_M}{r_C}\right)^2 = 0.605,\,0.593\,
\end{equation}
for the pion and proton, respectively. It is imperative, nonetheless, to compare and contrast the ratio $(r_M/r_C)^2$ across the literature. A representative sample is shown in Fig.\,\ref{radius} and discussed below. As indicated in Ref.~\cite{Broniowski:2008hx} and argued therein, if only the contributions of intrinsic quarks in chiral quark models are considered, the ratio  $(r_M/r_C)^2$ for pion is strictly $1/2$. On the other hand, it has been shown that factorized models of the pion light-front wave function establish a close relation between $(r_M/r_C)^2$ and the moments of the hadronic scale PDF~\cite{Raya:2021zrz}, entailing that this ratio is always a positive number less than unity. If one also considers the so-called \emph{physical bounds} of the corresponding PDF (see \emph{e.g.}\,\cite{Cui:2022bxn,Lu:2023yna}), $1/2$ appears as a lower bound stemming form a constant distribution function\,, so restricting the charge-mass ratio as follows:
\begin{equation}
    \frac{1}{2} \leq \left(\frac{r_M}{r_C}\right)^2 \leq 1 \,.\label{eq:lims}
\end{equation}
With such a premise, the data-driven analysis from \cite{Xu:2023bwv} yields $(r_M/r_C)^2=0.62(5)$, a value that falls within these bounds. Interestingly, the $(r_M/r_{C})^2=1$ limit would be compatible with the physical intuition that as the hadrons become heavier, charge and mass densities become more similar to each other, approaching the classical expectations\,\cite{Raya:2024ejx}. 

From an entirely different perspective, Ref.~\cite{Xing:2022mvk} highlights the importance of relevant meson poles in a consistent calculation of the pion EFF and GFFs. For instance, while it succeeds in naturally incorporating a vector meson pole in the electromagnetic form factor, the tensor pole is missing in the mass GFF, resulting in a larger charge radius whose size is not compensated in the mass radius, hence yielding $(r_M/r_{C})^2\lesssim 1/2$. On the other hand, a subsequent examination using a more sophisticated computational framework with some modeling elements~\cite{Xu:2023izo}, incorporates both vector and tensor poles in the corresponding form factors, obtaining a ratio that falls within the bounds given in Eq.\,\eqref{eq:lims}. The significance of meson poles is further discussed in the Ref.~\cite{Broniowski:2024oyk}, where considering solely the relevant poles better fits the lattice results, once again indicating $(r_M/r_{C})^2\leq 1$.

Finally, despite the consensus that the mass radius is indeed smaller than the electromagnetic radius, which our calculation also reveals for the proton, a detailed scrutiny of the $1/2$ lower limit in Eq.\,\eqref{eq:lims} would be profoundly  significant from an empirical perspective. We look forward to higher precision experiments and lQCD calculations to resolve this matter.

\section{Summary}\label{s4}
Based upon the LFHQCD approach, we have proposed a method to determine the GPD at zero skewness from its corresponding forward limit, \emph{i.e.}, the PDF. The study focuses on a simultaneous description of the pion and the proton, for which robust predictions of their corresponding PDFs are available from CSMs calculations\,\cite{Ding:2019qlr,Chang:2022jri}. It is noteworthy that various techniques have been developed to infer the hadronic scale PDF from either experimental data or lattice QCD moments  (\emph{e.g} Refs.\,\cite{Xu:2023bwv,Lu:2023yna,Wang:2023bmk,Cui:2021mom}), which could potentially eliminate this 
apparent bias observed in current approaches.

By construction, the EFFs derived within this framework exhibit remarkable behavior and conform to expected asymptotic limits\,\cite{Chang:2020kjj,deTeramond:2018ecg}, while showing insensitivity to the specific details of the PDF. This does not extend to the GPD itself or the GFFs, which are entirely predictions of the model. Our particular focus lies on the GFF $A(t)$, which is directly related to the mass distribution within the hadrons. The reason is largely technical, as it results directly from the calculation of the first moment of the GPD in the limit of zero skewness. However, as we have seen, it is sufficient to gain significant insights on the partonic structure of the hadrons. 

As a result of our investigation, we find very favorable agreement between our theoretical predictions for the (mass) gravitational form factors of the pion and proton and recent lattice QCD results in Refs~\cite{Hackett:2023nkr,Hackett:2023rif}, which were calculated close to the physical mass of the pion. The high-precision lattice QCD data on GFFs represent a milestone in our quest to understand the internal dynamics of hadrons, so the agreement cannot go unnoticed. Our analysis not only validates the ideas behind the identification of a hadronic scale, but also uncovers intriguing aspects of the mass and charge distributions. For instance, in both the pion and proton, the charge effects extend over a larger area. Additionally, as anticipated, the proton occupies a greater volume. While many theoretical and phenomenological analyses agree on these observations, the ratio of the associated $(r_M/r_C)^2$, in the case of the pion, varies over a somewhat generous range. In particular, our calculation for the pion places us within the so-called physical bounds in Eq.\,\eqref{eq:lims}, while both lattice QCD\,\cite{Hackett:2023nkr} and the phenomenological extraction\,\cite{Kumano:2017lhr} lie below it. In either case, this discrepancy could be attributed to subtleties of a technical or systematic nature. It is hoped that this issue will be resolved with improved precision in future extractions.

Looking ahead, we aim to continue exploring these topics, particularly extending our investigations to inquire on spin-related form factors and parton distributions. This includes the GFF $J(t)$ of the proton, which could be addressed in an analogous way via the corresponding helicity-flip GPD. Continuous advances in experimental techniques and theoretical approaches, such as LFHQCD and CSMs, hold promise for providing deeper insights into the most challenging aspects of QCD.

\section{Acknowledgement}
We are grateful to the authors of Refs.~\cite{Hackett:2023nkr,Hackett:2023rif} for kindly providing us the data tables used in Fig.~\ref{AFF}. This work is supported by the National Natural Science Foundation of China (grant no. 12135007), the Spanish MICINN grant PID2022-140440NB-C22, the
regional Andalusian project P18-FR-5057, and the Helmholtz-Zentrum Dresden-Rossendorf High Potential Programme.

\bibliography{bibreferences}



\end{document}